\documentclass[12pt]{iopart}
\usepackage[dvips]{graphicx}
\usepackage{graphicx}
\usepackage{subfig}
\usepackage{pdfpages}
\usepackage{pbox}
\usepackage{xcolor}
\usepackage{cite}
\usepackage[colorlinks, citecolor = blue, urlcolor = blue]{hyperref}
\begin{document}

\title[Electron-phonon superconductivity in C-doped topological nodal-line semimetal Zr$_5$Pt$_3$.....]{Electron-phonon superconductivity in C-doped topological nodal-line semimetal Zr$_5$Pt$_3$: A muon spin rotation and relaxation ($\mu$SR) study}

\author{A Bhattacharyya$^{1^*}$, P P Ferreira$^{2^{\dagger}}$, K Panda$^{1}$, F B Santos$^{2}$, D T Adroja$^{3,4}$, K Yokoyama$^{3}$, T T Dorini$^{5}$, L T F Eleno$^{2}$ and A J S Machado$^2$}
\address{$^{1}$ \scriptsize Department of Physics, Ramakrishna Mission Vivekananda Educational and Research Institute, Howrah 711202, India,
Howrah 711202, West Bengal, India}
\address{$^{2}$ \scriptsize Escola de Engenharia de Lorena, Universidade de São Paulo, DEMAR, Lorena, Brazil}
\address{\small$^{3}$ \scriptsize ISIS Facility, Rutherford Appleton Laboratory, Chilton, Didcot, Oxon, OX11 0QX, United Kingdom}
\address{\small$^{4}$ \scriptsize Highly Correlated Matter Research Group, Physics Department, University of Johannesburg, Auckland Park 2006, South Africa}
\address{\small$^{5}$ \scriptsize Université de Lorraine, CNRS, IJL, Nancy, France}

\ead{$^*$amitava.bhattacharyya@rkmvu.ac.in,\\~~~~~~ $^{\dagger}$pedroferreira@usp.br}

\begin{abstract}

\noindent In the present work we demonstrate that C-doped Zr$_{5}$Pt$_{3}$ is an electron-phonon superconductor (with critical temperature T$_\mathrm{C}$ = 3.7\,K) with a nonsymmorphic topological Dirac nodal-line semimetal state, which we report here for the first time. The superconducting properties of Zr$_{5}$Pt$_{3}$C$_{0.5}$ have been investigated by means of magnetization and muon spin rotation and relaxation ($\mu$SR) measurements. We find that at low temperatures the depolarization rate is almost constant and can be well described by a single-band $s-$wave model with a superconducting gap of $2\Delta(0)/k_\mathrm{B}T_\mathrm{C}$ = 3.84, close to the value of BCS theory. From transverse field $\mu$SR analysis we estimate the London penetration depth $\lambda_{L}$ = 469 nm, superconducting carrier density $n_{s}$ = 2$\times$10$^{26}$ $m^{-3}$, and effective mass m$^{*}$ = 1.584 $m_{e}$. Zero field $\mu$SR confirms the absence of any spontaneous magnetic moment in the superconducting ground state. To gain additional insights into the electronic ground state of C-doped Zr$_5$Pt$_3$, we have also performed first-principles calculations within the framework of density functional theory (DFT). The observed homogenous electronic character of the Fermi surface as well as the mutual decrease of $T_\mathrm{C}$ and density of states at the Fermi level are consistent with the experimental findings. However, the band structure reveals the presence of robust, gapless fourfold-degenarate nodal lines protected by $6_{3}$ screw rotations and glide mirror planes. Therefore, Zr$_5$Pt$_3$ represents a novel, unprecedented condensed matter system to investigate the intricate interplay between superconductivity and topology. 
    
\end{abstract}


\noindent 
\maketitle

\section{Introduction}

\noindent The search for new superconductors is a cornerstone in quantum matter physics. In particular, the study of how the superconducting state evolves with doping and pressure has been essential to understand the fundamental mechanisms of Cooper pairs condensation and competion between different coherent states. However, in the last few years we are seeing a new chapter being written, where the topological nature of matter is responsible for the emergence of a wide range of novel quantum states. Topological insulators and symmetry-protected topological semimetals can exhibit quantizied anomalous Hall effect \cite{haldane2004,xu2011,weng2015}, ultrahigh electronic mobility \cite{shekhar2015,liang2015,zhao2015}, negative/giant magnetoresistence \cite{li2016,gao2017}, chiral anomaly \cite{zyuzin2012,parameswaran2014,xiong2015,huang2015}, and gapless, robust edge states \cite{hsieh2008,hsieh2009,zhang2009}, to cite a few examples. When combined with superconductors, topological materials could also harbor Majorana quasiparticles with non-Abelian exchange statistics, thus possibly opening the way for the quantum computation era \cite{beenakker2013}.

\noindent In this context, a large number of intermetallic compounds with general formula M$_{5}$X$_{3}$ (M is either a transition or rare earth metal and X is a metalloid) represent a novel platform to explore a plethora of unique properties~\cite{surgers2003, lv2013, Xu2020, Zhang2017, Zheng2002, Cort1982, Bortolozo2012, Hamamoto2018, Renosto2018, Claeson}. These compounds can crystallize in three different prototypes: (i) orthorhombic Yb$_\mathrm{5}$Sb$_\mathrm{3}$ ($Pnma$, No.\,62), (ii) tetragonal Cr$_{5}$B$_{3}$ ($I4/mcm$, No.\,140), and (iii) hexagonal Mn$_{5}$Si$_{3}$ ($P6_{3}/mcm$, No.\,193). Among them, the Mn$_\mathrm{5}$Si$_\mathrm{3}$ structure can interstitially host a third element (carbon, boron, nitrogen or oxygen) at 2$b$ Wyckoff position. Physical properties of more than 500 compounds in this hexagonal structure have been reported so far. Superconductivity is observed only in few  compounds though, such as Zr$_{5}$Sb$_{3}$ (T$_\mathrm{C}$ = 2.3\,K)~\cite{lv2013}, Zr$_{5}$Pt$_{3}$ (T$_\mathrm{C}$ = 6.4 K)~\cite{Hamamoto2018}, tetragonal (T$_\mathrm{C}$ = 2.8\,K) and hexagonal (T$_\mathrm{C}$ = 9.4\,K) Nb$_{5}$Ir$_{3}$~\cite{Cort1982}, and Nb$_{5}$Ge$_{3}$ (T$_\mathrm{C}$ = 0.3 K)~\cite{Claeson}. In Nb$_{5}$Ir$_{3}$, for instance, T$_\mathrm{C}$ increases to 10.5\,K with oxygen doping~\cite{Zhang2017} and a crossover from multiple to single gap superconductivity with increasing Pt content was reported in Nb$_{5}$Ir$_{3-x}$Pt$_{x}$O~\cite{Xu2020}. On the other hand, the addition of oxygen in Zr$_{5}$Pt$_{3}$ reduces monotonically the critical temperature from 6.4\,K to 3.2\,K (Zr$_{5}$Pt$_{3}$O$_{0.6 })$~\cite{Hamamoto2018}. Likewise, T$_\mathrm{C}$ of Zr$_{5}$Sb$_{3}$ decreases with the addition of oxygen until the supresison of the superconducting phase in Zr$_{5}$Sb$_{3}$O~\cite{lv2013}. The highest reported T$_\mathrm{C}$ in this family, however, belongs to Nb$_{5}$Ge$_{3}$C$_{0.3}$ \cite{Bortolozo2012}, with T$_{\mathrm{C}} = 15.3$\,K. 

 \begin{figure*}[t]
\centering
   \includegraphics[height=0.45\linewidth,width=0.9\linewidth]{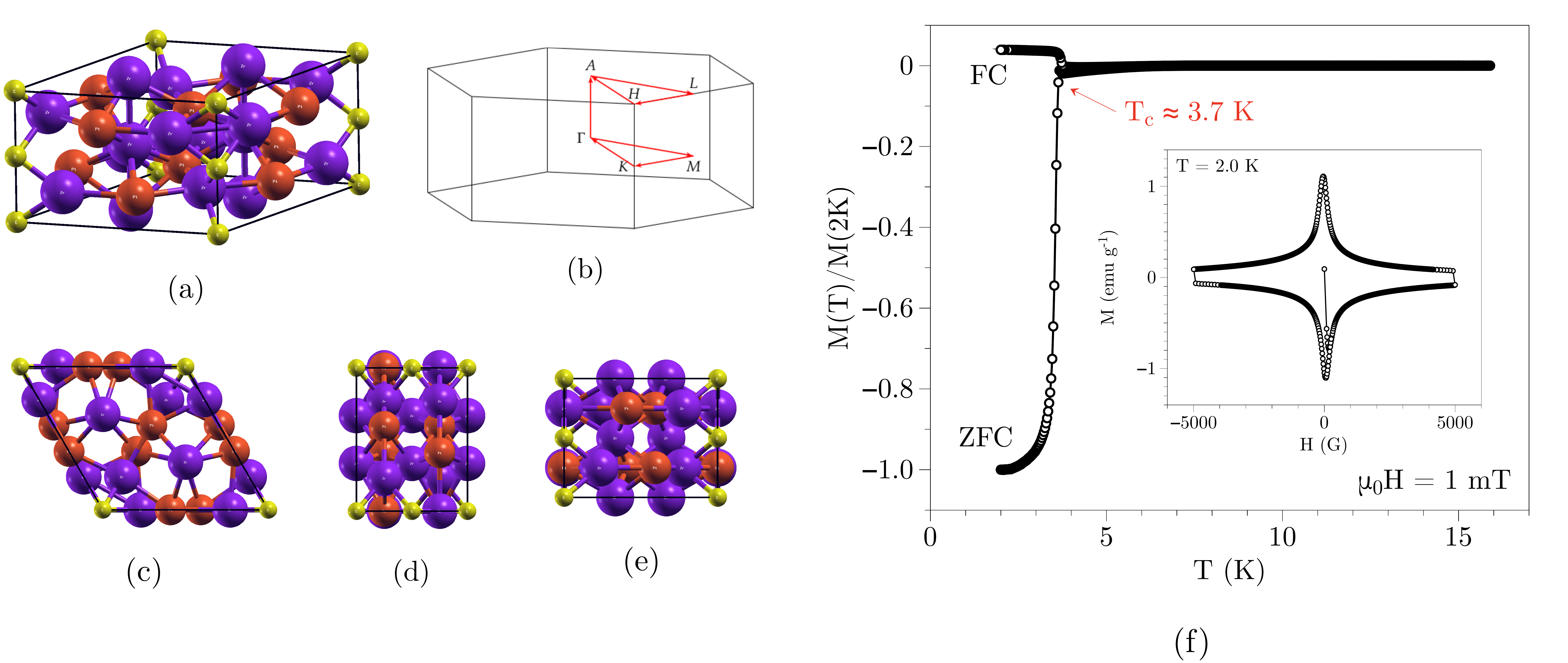}
\caption{(a) Hexagonal Zr$_{5}$Pt$_{3}$C  unit cell with space group $P6_{3}/mcm$ (no. 193). Purple, orange, and yellow spheres represent Zr, Pt, and C, respectively. (b) First Brillouin zone of Zr$_5$Pt$_3$ system with the path along high-symmetry points. The $a$--$b$, $a$--$c$, and $c$--$b$ planes are shown in (c), (d) and (e), respectively. (f) Temperature variation of the ac magnetic susceptibility in presence of an applied field of 1\,mT, collected in FC and ZFC modes. The inset shows the field dependence of the isothermal magnetization at 2 K.}
\label{physical properties}
\end{figure*}

\noindent Recently, Renosto {\it et al.} reported a comprehensive study of the effects of carbon doping on the superconducting properties of Zr$_{5}$Pt$_{3}$C$_x$ \cite{Renosto2018}. It was found that T$_\mathrm{C}$ first increases from 6\,K ($x=0$) to 7\,K ($x=0.3$) and further decreases monotonously for $x > 0.3$~\cite{Renosto2018}. However, by measuring the lower and upper critical fields and the temperature dependence of penetration lenght and specific heat, the authors argued that Zr$_5$Pt$_3$C$_{0.3}$ strongly deviates from the conventional s-wave pairing behavior, suggesting thus an unconventional superconductivity driven by carbon doping.

\noindent Motivated by these results, we show, using muon spin relaxation and rotation ($\mu$SR) measurements combined with first-principles electronic-structure calculations, that a conventional, single-gap s-wave superconducting order parameter within the weak-coupling limit is sufficient to describe the superconducting ground state of Zr$_{5}$Pt$_{3}$C$_{0.5}$, in contrast to the recent proposals. This is  the first systematic $\mu$SR study on the superconducting properties of doped M$_\mathrm{5}$X$_\mathrm{3}$-type compounds. Furthermore, we demonstrate that Zr$_{5}$Pt$_3$ is a topological nodal-line semimetal protected by nonsymmorphic symmetries. In contrast to Dirac/Weyl semimetals, where the conduction and valence bands touch at discrete points in the Brillouin zone and disperse linearly in all momentum directions \cite{bernevig2018}, in nodal-line semimetals the linear band-crossing points form closed loops in momentum space, opening the way to novel quantum phases, such as nearly flat drumhead-like surface states \cite{weng2015-2,yu2015,chan2016}. In this vein, Zr$_5$Pt$_3$-type compounds represent an unprecedented, large family of superconducting nodal-line semimetals, still to be explored in more details.

\section{Experimental Details}

\subsection{Sample Preparation and Physical Properties Measurements}

\noindent A Zr$_{5}$Pt$_{3}$C$_{0.5}$ sample was prepared using typical arc melting process in argon gas atmosphere on a water-cooled copper crucible using highly pure Zr, Pt, and C in a stoichiometric ratio. The arc melted ingot was remelted several times to confirm its homogeneity. After that, the sample was annealed at 1200\,$^\mathrm{o}$C for a week in a sealed vacuum quartz tube. X-ray powder diffraction was carried out using Cu-$K_{\alpha}$ radiation. The magnetization measurements were performed using VSM-SQUID, 9T-PPMS (Quantum Design Inc.) using both field-cooled (FC) and zero-field-cooled (ZFC) protocols. 

\subsection{$\mu$SR Measurements}

\noindent $\mu$SR measurements were carried out at the ISIS Neutron and Muon Source of  STFC Rutherford Appleton Laboratory, UK, on the MUSR spectrometer. 64 detectors are placed in transverse and longitudinal directions to detect the asymmetry of the positrons~\cite{Lee1999}. In the sample, 100\% spin-polarized positive muons are implanted, which decay with a half-life of 2.2$\mu$s, decaying into a positron and a pair of neutrinos. As the positrons are preferentially emitted along the muon spin vectors, information on the local magnetic field distribution at the muon stopping site can be obtained by detecting the asymmetry of the emitted positrons. The time-dependent symmetry $A(t)$ of the $\mu$SR spectra is given by $A(t)=\frac{N_{\mathrm{F}}(t)-\alpha N_{\mathrm{B}}(t)}{N_{\mathrm{F}}(t)+\alpha N_{\mathrm{B}}(t)}$, where $N_{\mathrm{F}}(t)$ and $N_{\mathrm{B}}(t)$ are the number of positrons counted in the forward and  backward detectors respectively, and $\alpha$ is an instrumental calibration factor. ZF-$\mu$SR is carried out in the longitudinal set up of the detectors. A correction coil is applied to neutralize any stray magnetic fields up to 10$^{-3}$~G. The transverse field measurements were carried with detectors in a transverse arrangement, with a field of 300 G ( well below the upper critical field $H_\mathrm{c2}$=6.3 T) applied perpendicular to the initial muon polarization direction~\cite{ISIS2018}.

\noindent The Zr$_{5}$Pt$_{3}$C$_{0.5}$ sample was powdered and placed in a high purity (99.995\%) silver sample holder using diluted GE-varnish and then wrapped with thin silver foil, since the signal from muons stopping in silver depolarizes at a negligible rate. All data analysis were done using WiMDA~\cite{Pratt2000} software. 
   
\subsection{Computational Methods}

\noindent First-principes electronic-structure calculations were performed within the Kohn-Sham scheme \cite{kohn1965} of the Density Functional Theory (DFT) \cite{hohenberg1964} with full-relativistic projector augmented plane wave pseudopotentials \cite{dal2014} as implemented in Quantum \textsc{Espresso} \cite{giannozzi2009,giannozzi2017}. Exchange and correlation (XC) effects were treated with the local density approximation (LDA) as described by the Perdew-Zunger (PZ) parametrization \cite{perdew1981}. We have used a wave function energy cut-off of 80\,Ry (1\,Ry $\approx$ 13.6\,eV), and 800\,Ry for the charge density and potential kinetic energy cut-off. The Monkhorst-Pack scheme \cite{monkhorst1976} was used for a $8\times8\times12$ $k$-point sampling in the first Brillouin zone. A denser $16\times16\times24$ $k$-point sampling was further used to obtain the band structure, density of states and Fermi surface. Self-consistent-field (SCF) calculations were carried out using a Marzari-Vanderbilt smearing \cite{marzari1999} of 0.005\,Ry. All latice parameters and internal degrees of freedom were relaxed in order to guarantee a ground-state convergence of 10$^{-5}$\,Ry in total energy and 0.5\,mRy/$a_0$ ($a_0\approx0.529\,$\AA) for forces acting on the nuclei.

\section{Results and discussion}

\subsection{Crystal structure and Magnetization}

\noindent X-ray powder diffraction shows that Zr$_5$Pt$_3$ crystallizes in hexagonal Mn$_{5}$Si$_{3}$-type structure ($P6_3/mcm$, No.\,193), as shown in Fig.~\ref{physical properties}(a).  The temperature dependence of the magnetic susceptibility, $\chi(T)$, in the presence of an applied field of 10 G is presented in Fig.~\ref{physical properties}(b). The low field dc magnetic susceptibility clearly shows a superconducting phase transition at T$_\mathrm{C} = 3.7$\,K. Moreover, the isothermal field dependence of magnetization at 2\,K [see the inset in Fig.~\ref{physical properties}(b)] confirms the presence of a type-II superconductivity.  
 
\begin{figure}[t]
\centering
    \includegraphics[width=0.5\linewidth]{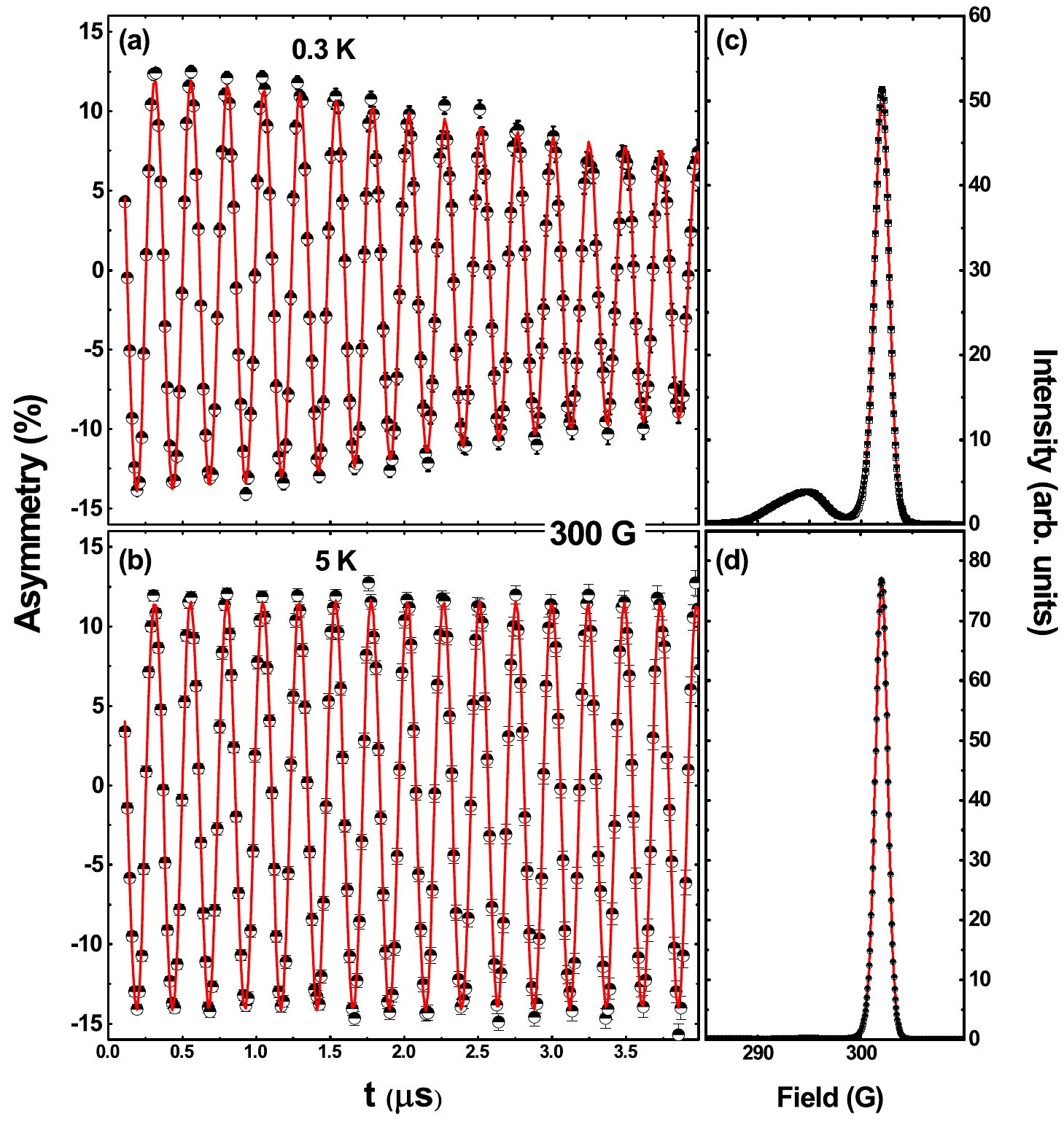}
    \caption{(a) Time dependence of transverse field $\mu$SR asymmetry spectra collected at (a) $T$ = 0.3 K ($<$T$_\mathrm{C}$) and (b) $T$ = 5 K ($>$T$_\mathrm{C}$) in the presence of an applied field $H$ = 300 G. The red solid line shows the fit to the data using Eq.~(\ref{eq:fit}). (c) and (d) display the corresponding maximum entropy spectra (below and above T$_\mathrm{C}$). }
\label{musrdata1}
\end{figure}

\subsection{TF-$\mu$SR analysis}

\begin{figure}[t]
\centering
    \includegraphics[width=0.6\linewidth]{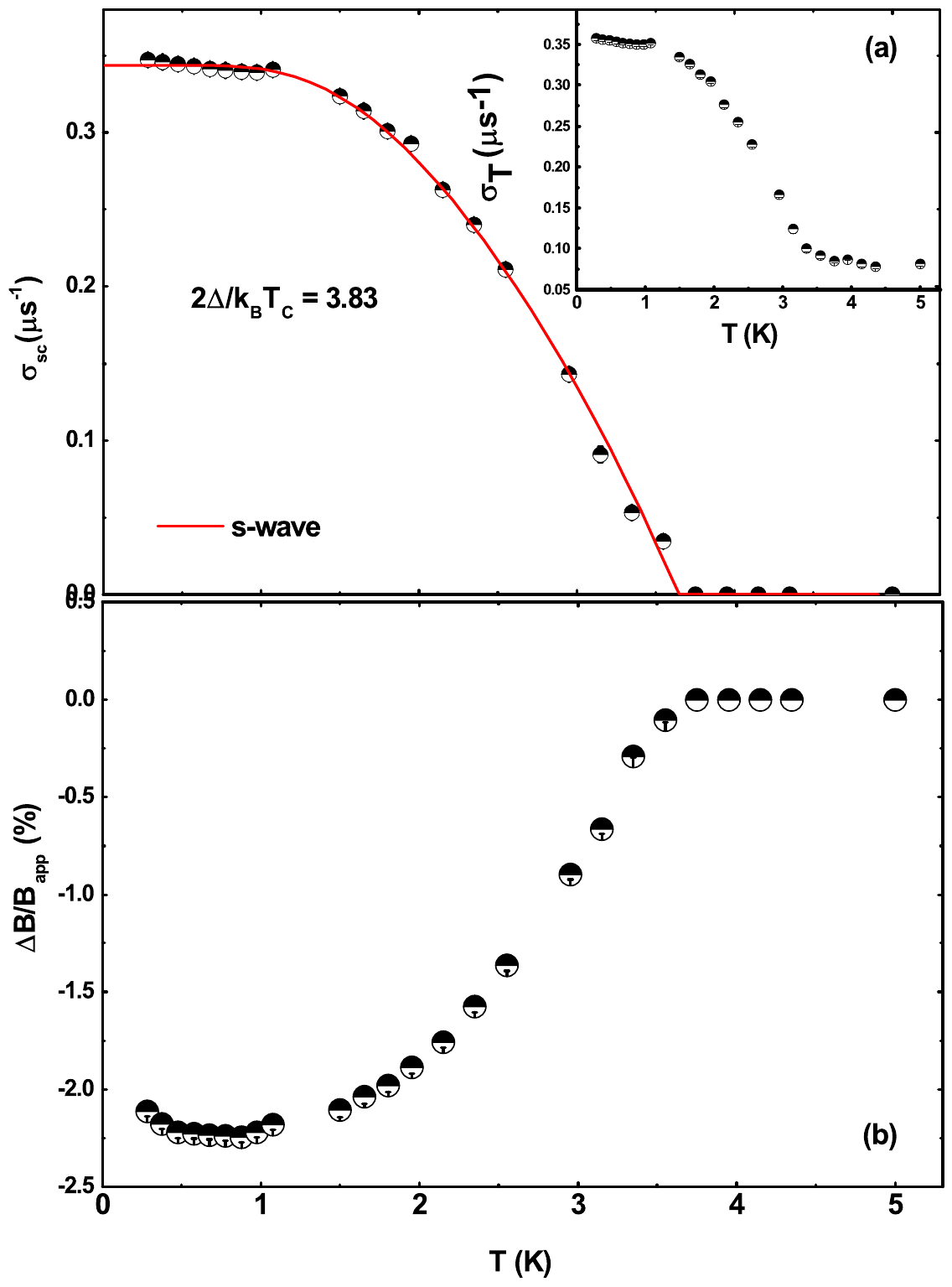}
    \caption{(a) The temperature variation of superconducting depolarization rate $\sigma_\mathrm{sc}(T)$. The solid red line shows the fit using $s-$wave model. Inset shows the total muon spin depolarization rate $\sigma_\mathrm{T}$ as a function of temperature.  (b) The relative change of the internal field normalized to external applied field as a function of temperature, where $\Delta B$ = B$_\mathrm{SC}$ - B$_\mathrm{app}$.}
\label{musrdata2}
\end{figure}

\noindent TF-$\mu$SR measurements  were performed from 0.3\,K to 5\,K. The observed asymmetries, at 0.3\,K and 5\,K, below and above $T_\mathrm{C}$ respectively, can be found in Fig~\ref{musrdata1}(a)-(b). The respective maximum entropies are shown in Fig.~\ref{musrdata1}(c) and (d). In the superconducting state at 0.3\,K, it is clear that the spectra strongly depolarize due to the inhomogeneous distribution of the internal field derived from the formation of the vortex state. Meanwhile, at 5\,K the spectra depolarization is quite negligible, which can be attributed to the nuclear moments of the silver background. The time spectra can be fitted using two Gaussian oscillatory functions \cite{bhattacharyya2018,adroja2017,bhattacharyya2019},
\begin{equation}
G_\mathrm{TF}(t) = A_\mathrm{sc}\cos(\omega_\mathrm{sc}t+\phi)\exp(-\frac{\sigma_\mathrm{T}^{2}t^{2}}{2})+A_\mathrm{bg}\cos(\omega_\mathrm{bg}t+\phi),
\label{eq:fit}
\end{equation}
\noindent where the initial asymmetries associated with both sample and background are represented by $A_{\mathrm{sc}}$ and $A_{ \mathrm{bg}}$, respectively, $\omega_\mathrm{sc}$ and $\omega_\mathrm{bg}$ are the frequencies of the muon spin precession frequency from the sample and background, and $\phi$ is the phase offset. The Gaussian depolarization rate $\sigma_{T}$ has two contribution below $T_\mathrm{C}$, where $\sigma_\mathrm{sc}$ is derived from a superconducting component and $\sigma_\mathrm{nm}$(=0.0852 $\mu$s$^{-1}$) is derived from a nuclear magnetic dipolar moment that is constant over the whole temperature range, later assisted by the ZF-$\mu$SR. Thus, the contribution from the vortex lattice, $\sigma_{sc}$, was calculated by quadratically subtracting $\sigma_\mathrm{nm}$ obtained from the fitting of the spectra measured above $T_\mathrm{C}$. The field shift is $\Delta B$ = B$_\mathrm{SC}$-B$_\mathrm{app}$, where B$_\mathrm{SC}$ indicates the superconducting field induced by the vortex lattice and B$_\mathrm{app}$ is the applied field, as shown in Fig. \ref{musrdata2}(b). As the sample goes through the transition into the superconducting state, there is a strong negative shift in the peak field, which is a unique characteristic of the vortex lattice \cite{brandt1988,brandt2003}.

\noindent Figure~\ref{musrdata2}(a) presents $\sigma_\mathrm{sc}(T)$ as a function of temperature, which is proportional to the superfluid density, thus providing details about the gap structure. It is clear that at low temperature the $\sigma_\mathrm{sc}$ is almost constant, which indicates a fully gapped superconducting state. The normalized superfluid density was modelled using~\cite{prozorov2006, adroja2015, adroja2017-2, HfIrSi, bhattacharyya2021}
\begin{eqnarray}
\frac{\sigma_{sc}(T)}{\sigma_{sc}(0)} &=& \frac{\lambda^{-2}(T)}{\lambda^{-2}(0)}\\
 &=& 1 + \frac{1}{\pi}\int_{0}^{2\pi}\int_{\Delta(T)}^{\infty}(\frac{\delta f}{\delta E}) \times \frac{EdEd\phi}{\sqrt{E^{2}-\Delta(T})^2} \nonumber.
\end{eqnarray}
\noindent Here, $f$ is the Fermi function, that can be represented by $f= [1+\exp(-E/k_\mathrm{B}T)]^{-1}$, and $\Delta(T,0) = \Delta_{0}\delta(T/T_\mathrm{C})g(\phi)$, whereas $g(\phi)$ is the angular dependence of the gap function. The azimuthal angle in the direction of the Fermi surface is denoted by $\phi$. The temperature variation of the superconducting gap is approximated by the relation $\delta(T/T_\mathrm{C}) = \tanh \{{1.82[1.018 (T_\mathrm{C}/T -1)]^{0.51}} \}$. $g(\phi$), the spatial dependence, is substituted by 1 for $s-$wave symmetry. Using this, we find that the data is best modelled considering a single isotropic $s$-wave gap of 0.59\,meV, which yields $2\Delta(0)/k_\mathrm{B}T_\mathrm{C} = 3.84$, close to the value of 3.53 predicted for weak-coupling BCS superconductors. Below T$\mathrm{C}$ the electronic specific heat capacity (C$_\mathrm{e}$) is well described by BCS gap model~\cite{Renosto2018}.

\noindent The depolarization rate ($\sigma_\mathrm{sc}$) recorded below $T_\mathrm{C}$ is correlated with the superfluid density or the penetration depth. For a triangular lattice,  $\frac{\sigma_\mathrm{sc}^2}{\gamma_\mathrm{\mu}^2}=\frac{0.00371 \times \phi_\mathrm{0}^{2}}{\lambda^4}$ \cite{Sonier,chia2004,amato1997}, where $\phi_{\mathrm{0}}$ is the flux quantum number (2.07 $\times$10$^{-15}$T\,m$^{2}$) and $\gamma_{\mathrm{\mu}}$ is the muon gyromagnetic ratio $\gamma_\mathrm{{\mu}}/2\pi$ = 135.5 MHz\,T$^{-1}$. Superfluid density can also be related to quantities at the atomic order, as with other phenomenological parameters that characterize the superconducting state. Using London's theory, \cite{Sonier} $\lambda_{\mathrm{L}}^2 = \frac{m^{*}c^{2}}{4\pi n_\mathrm{s}e^{2}}$, where $m^{*} = (1+\lambda_\mathrm{e-ph})m_\mathrm{e}$ is the effective mass and $n_\mathrm{s}$ is the density of superconducting carriers. Within this simple picture, $\lambda_{\mathrm{L}}$ is independent of magnetic field. $\lambda_{\mathrm{e-ph}}$ is the electron-phonon coupling parameter that can be estimated from the Debye temperature ($\Theta_{\mathrm{D}}$) and $T_{\mathrm{C}}$ using the semi-empirical McMillan equation \cite{mcmillan1968},
\begin{equation}
\lambda_\mathrm{e-ph} = \frac{1.04+\mu^{*}\ln(\Theta_\mathrm{D}/1.45T_\mathrm{C})}{(1-0.62\mu^{*})\ln(\Theta_\mathrm{D}/1.45T_\mathrm{C})-1.04},
\end{equation}
where $\mu^{*}$ is a repulsive electron-electron pseudopotential with typical values at the order of $\mu^{*}$ = 0.1, which gives $\lambda_{\mathrm{e-ph}} = 0.584$. Since Zr${_5}$Pt${_3}$C${_{0.5}}$ is a type-II superconductor, it implies that the density of the normal state carriers is approximately equal to the density of superconducting carriers ($n_\mathrm{s}\approx n_\mathrm{e}$). Therefore, the density of the superconducting carriers ($n_{s}$), their effective-mass ($m^{*}$), and London penetration depth ($\lambda_{L}$) can be estimated, respectively, as $m^{*} = 1.584m_\mathrm{e}$, $n_\mathrm{s} = 2.026 \times 10^{26}$ carriers/m$^{3}$, $\lambda_\mathrm{L}(0)$ = 469~nm, for Zr$_{5}$Pt$_{3}$C$_{0.5}$.

\begin{figure}[t]
\centering
    \includegraphics[width=0.9\linewidth]{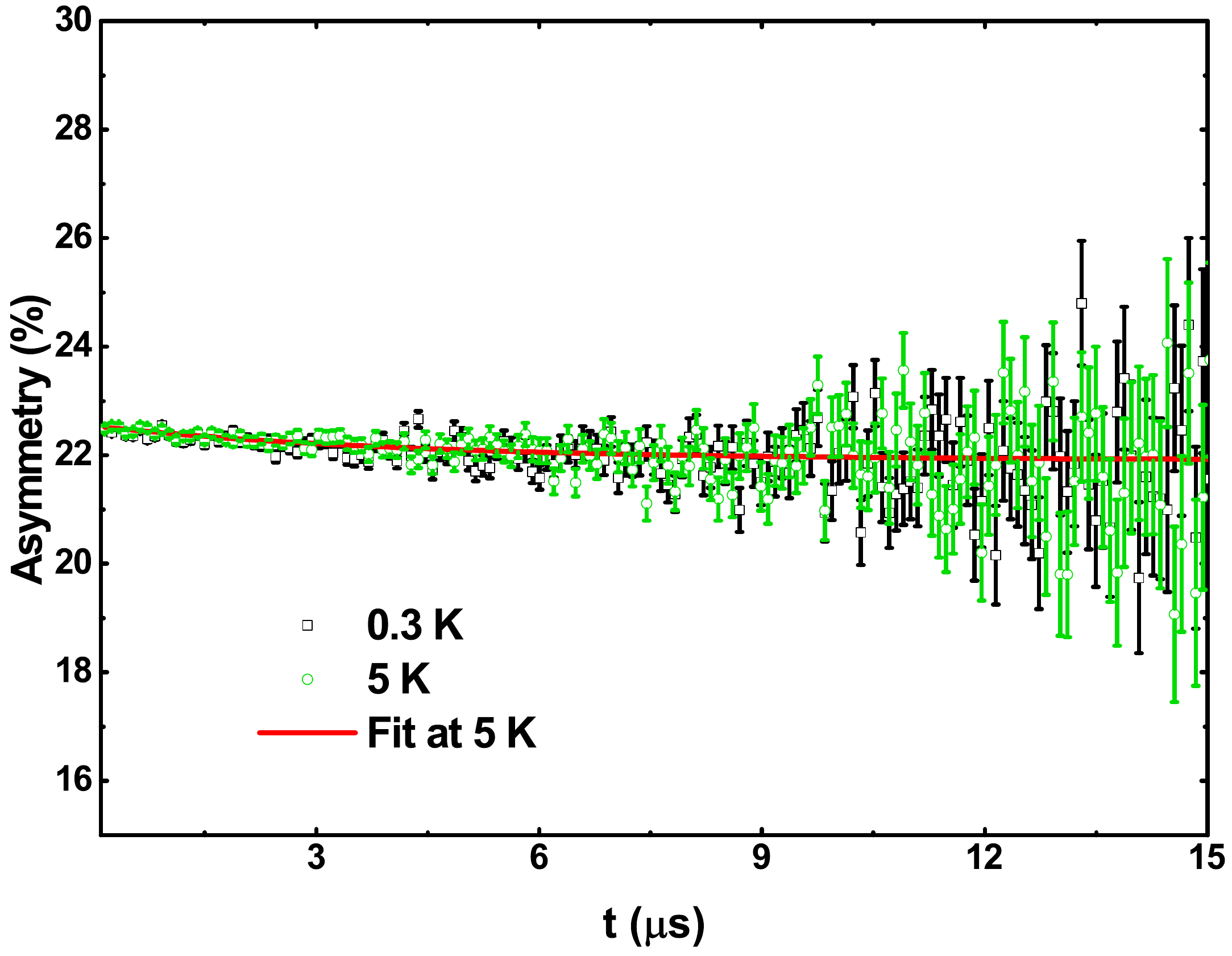}
\caption{Time evolution of ZF-$\mu$SR asymmetry spectra at  0.3 K (black squares) and 5 K (green circles), presented together. The red line is the least squares fit to the data using Eq.~(\ref{GZF}).}
\label{ZF}
\end{figure}

\subsection{ZF-$\mu$SR analysis}

\noindent ZF-$\mu$SR is used to investigate the existence of any spontaneous magnetic moment in the superconducting ground state \cite{bhattacharyya2015, bhattacharyya2015-2, bhattacharyya2018-2}. The evolution of the ZF-asymmetry spectra over time is displayed in Fig.~\ref{ZF} for $T$ = 0.3 K ($< T_\mathrm{C}$) and $T$ = 5 K ($> T_\mathrm{C}$). The spectra below and above $T_\mathrm{C}$ are found to be similar, ruling out the presence of any magnetic ordering, which implies that time-reversal symmetry (TRS) is preserved in the superconducting state of Zr$_{5}$Pt$_{3}$C$_{0.5}$. The ZF-$\mu$SR data was fitted using a Lorentzian function with a constant background \cite{panda2019}:
\begin{equation}
G_\mathrm{ZF}(t) = A_\mathrm{0}\exp{(-\lambda t)}+A_\mathrm{bg}.
\label{GZF}
\end{equation} 
Here, $A_\mathrm{0}$ is the sample asymmetry and $A_\mathrm{bg}$ is the background asymmetry, which are almost independent of the temperature. The parameter $\lambda$ is the rate of relaxation resulted from the nuclear moments. In Fig.~\ref{ZF}, the red line indicates the obtained fit for ZF-$\mu$SR data using Eq. \ref{GZF}. The parameters obtained from the ZF-$\mu$SR asymmetry data are as follows: $\lambda$ = 0.291 $\mu \mathrm{s}^{-1}$ at  0.3 K  and $\lambda$ = 0.242 $\mu \mathrm{s}^{-1}$ at 5 K.  The shift in the relaxation rate is within the error bar, indicating that TRS is preserved in Zr$_{5}$Pt$_{3}$C$_{0.5}$.

\section{Theoretical Calculations}

\begin{figure*}[t]
\begin{center}
	\subfloat[Zr$_5$Pt$_3$]{\includegraphics[width=.33\linewidth]{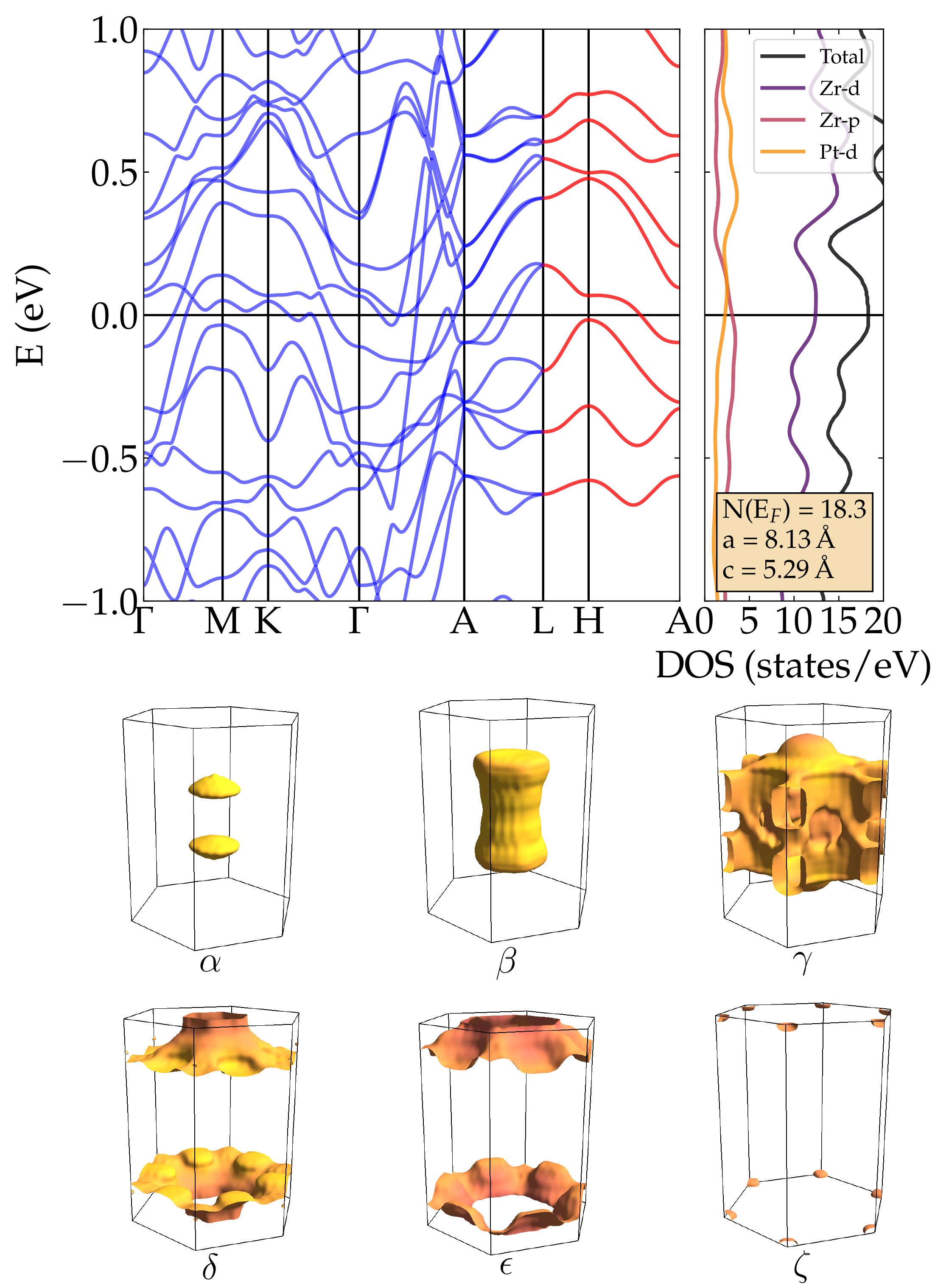}}
	\subfloat[Zr$_5$Pt$_3$C$_{0.5}$]{\includegraphics[width=.33\linewidth]{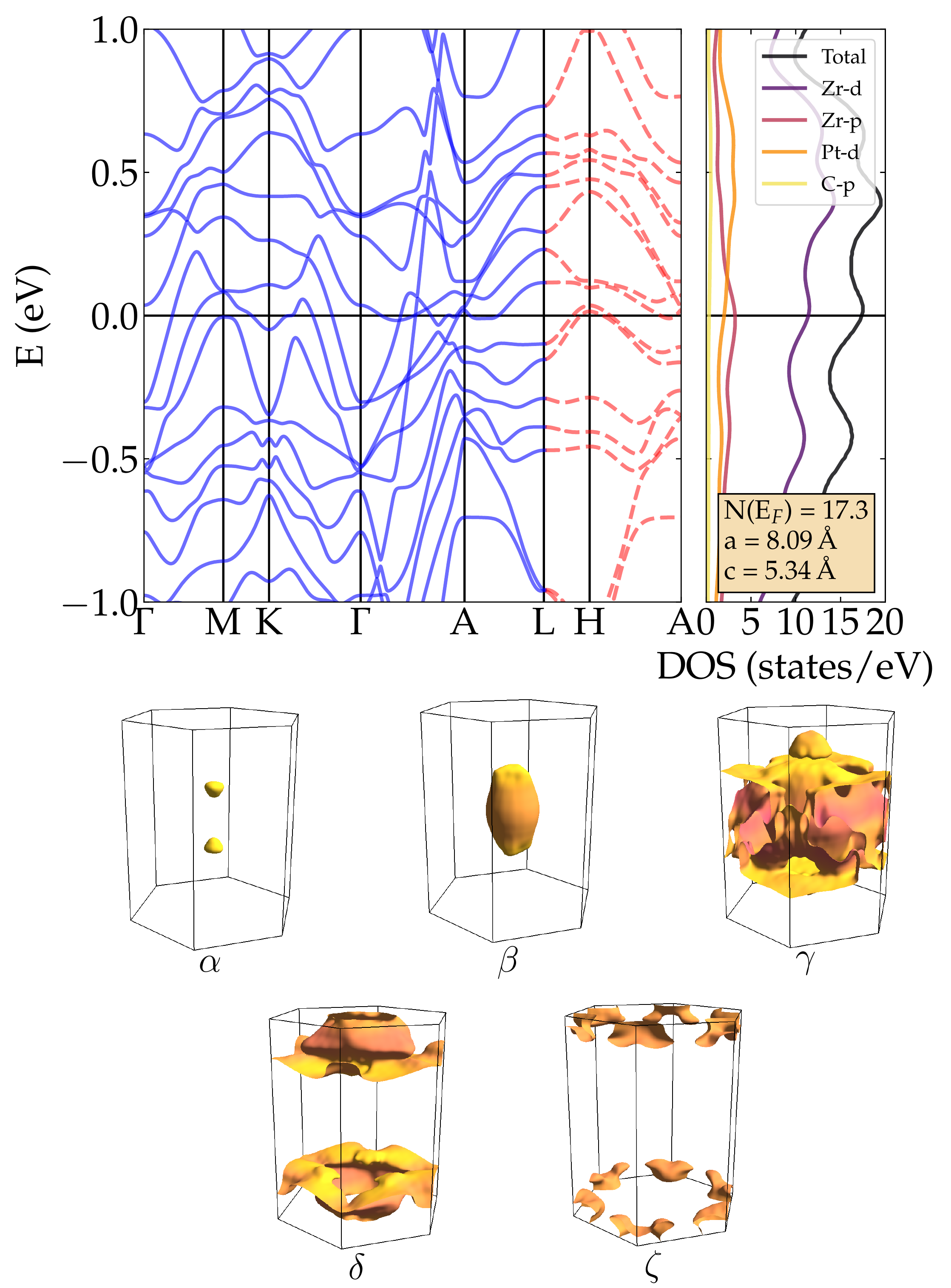}}
	\subfloat[Zr$_5$Pt$_3$C]{\includegraphics[width=.33\linewidth]{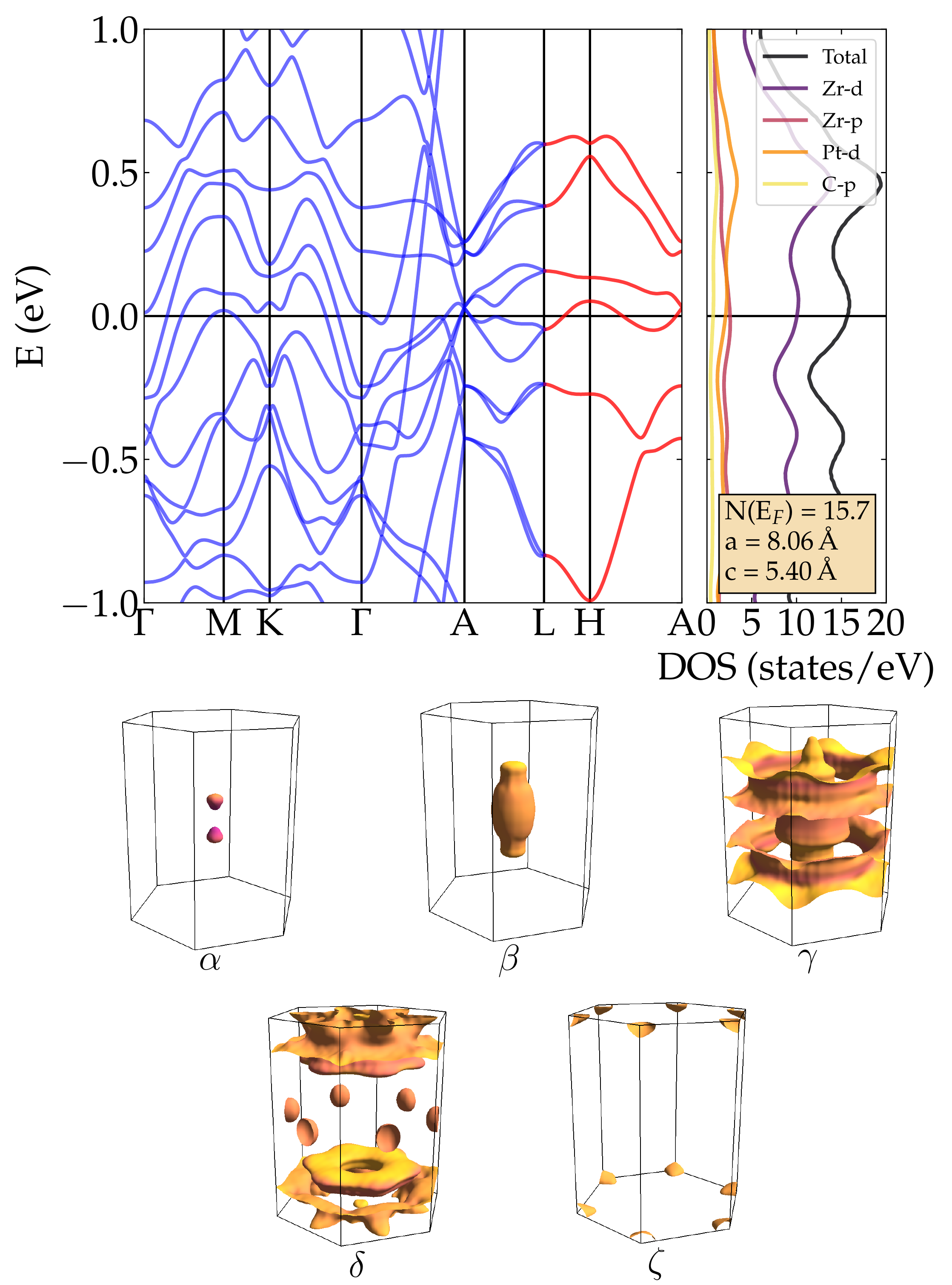}}\\		
	\includegraphics[width=.25\linewidth]{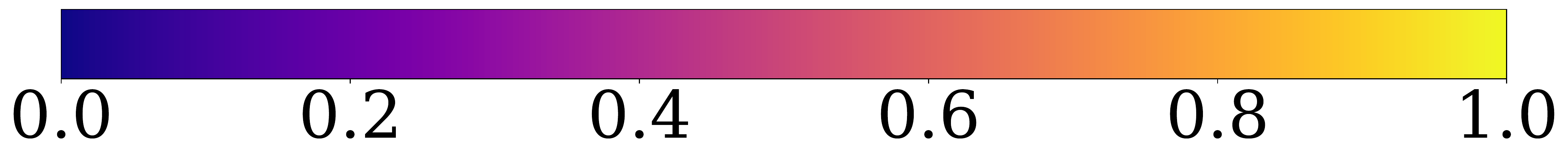}
	\caption{Electronic structure, partial density of states, and Fermi surface of (a) Zr$_5$Pt$_3$, (b) Zr$_5$Pt$_3$C$_{0.5}$, and (c) Zr$_5$Pt$_3$C. Also shown are the total DOS at the Fermi level, $N(E_F)$, and the relaxed lattice parameters. The gapless nodal lines protected by the nonsymmorphic 6$_3$ screw rotation symmetry of space group $P6_3/mcm$ are highlighted by solid red lines in the band structure diagram. The color map in the Fermi surfaces indicates the contribution of Zr-d orbitals in the low-energy states' wave functions.}
	\label{fig:bands}
	\end{center}
\end{figure*}

\noindent The electronic band structure, partial density of states (DOS), and Fermi surface of Zr$_{5}$Pt$_3$, Zr$_{5}$Pt$_3$C$_{0.5}$, and Zr$_{5}$Pt$_3$C including spin-orbit coupling (SOC) effects are shown in Figures \ref{fig:bands}(a)-(c), with the corresponding DOS at the Fermi level, $N(E_F)$, and optimized lattice parameters. With the increasing of C content $x$, the $a$ ($c$) lattice parameter decreases (increases) monotonically, reflecting a stronger in-plane hybridization between C-p and Zr-d states.  There are six bands crossing the Fermi energy in Zr$_5$Pt$_3$ [see Figure \ref{fig:bands}(a)], which results in a high density of states at the Fermi level of 18.3\,eV$^{-1}$, with approximately 68\,\% of these carriers derived from the Zr-d manifold. The change in $N(E_F)$ with increasing $x$ from 0 to 0.5 in Zr$_5$Pt$_3$C$_x$ is almost $-5.5\,$\%, which can be partily associated with the quasi-rigid band tuning of the chemical potential and the respective suppression of the $\epsilon$ sheet in the Fermi surface up to $x=1$. 

\noindent Although its complex, disconnected multiband nature, the homogeneous distribution of the electronic character of the Fermi surface shows an evident contrast from the usual signature presented by the Fermi surface of multiband superconductors, which generally presents very distinct orbital characters and an anisotropic hybridization on distinct sheets, and, as consequence, the average of the electron-phonon scattering connecting different points at Fermi surface are disjointed relative to the band index \cite{floris2007, bersier2009, flores2015, kawamura2017, ferreira2018, pascut2019, bhattacharyya2020, zhao2020,correa2021}. The mutual monotonic decrease of the DOS at the Fermi level and the observed superconducting critical temperature with the carbon doping \cite{Renosto2018} is compatible with the BCS theory as well. Therefore, in the light of our theoretical and experimental findings, the single-band electron-phonon s-wave pairing within weak coupling limit may be considered a coherent mechanism for superconductivity in Zr$_5$Pt$_3$ system. 

\noindent Interestingly, our calculations reveal that Zr$_5$Pt$_3$ is a nonsymmorphic topological Dirac nodal-line semimetal \cite{burkov2011, fang2015, fang2016, yu2017}, as explained below. In the presence of SOC all bands are twofold-degenerate due to the presence of both time-reversal and inversion symmetries. Along $\Gamma$--$A$ we can observe the presence of several linear band crossings in the vicinity of the Fermi level. Such fourfold-degenerate gapless Dirac nodes along $\Gamma$--$A$ are protected against SOC by $C_3$ rotational symmetry of the double point group $C_{6v}$, since both twofold-degenerate electronic dispersions that originate the band crossings, with distinct irreducible representations, possess different rotation eingevalues on the out-of-plane axis \cite{young2012, yang2014, yang2015, ferreira2021}. However,  pairs of twofold-degenerate bands along $A$--$L$ merge into single, degenerate bands along $L$--$H$--$A$, that is, along $k_z = \pi/c$ plane, which creates symmetry-enforced fourfold-degenerate nodal lines (solid red lines in Figure \ref{fig:bands}). These nodal lines are protected by 6$_3$ screw rotations with centre of symmetry in the 2$b$ Wyckoff position and axial glide mirror planes of the $P6_3/mcm$ space group, which means that they are robust against any pertubations as long as the nonsymmorphic symmetries hold. In fact, the presence of nonsymmorphic symmetries, such as screw rotations and glide mirror symmetries, support a nontrivial band connection at the Brillouin zone boundary \cite{yang2015}. 

\noindent The nonsymmorphic protection of the nodal lines are demonstrated in Figure \ref{fig:bands}(b), where we show the band structure of Zr$_5$Pt$_3$C$_{0.5}$. In this structure only half of the 2$b$ Wyckoff positions are filled, thus explicitly breaking the center of symmetry of the $6_3$ screw axis and the fractional translation of the glide mirror planes. Consequently, the nodal lines are gapped due to the spin-orbit coupling effect, as represented by the dashed red lines along $L$--$H$--$A$, and the Fermi surface acquire a distorsive topography. However, when the nonsymmorphic symmetries are preserved, by completely filling out the $2b$ positions in Zr$_5$Pt$_3$C [Figure \ref{fig:bands}(c)], the nodal-line states remain protected. At this point it is important to note that, on average, effective point-group symmetries can be preserved in a homogeneously, perfectly randomly disordered alloy, thus we expect that the topological nodal-line phase could be observed even upon a high level of alloying \cite{dziawa2012, narayan2014, yan2014, lu2017, thirupathaiah2018}, as in the case of Zr$_5$Pt$_3$C$_{0.5}$.  

\section{Conclusions}

\noindent In summary, we presented the superconducting properties of the interstitial carbon-doped compound Zr$_{5}$Pt$_{3}$C$_{0.5}$ using magnetization and muon spin rotation and relaxation measurements. Magnetization data confirms the bulk superconductivity at T$_{C}$ = 3.7 K. The depolarization rate of muon spin ($\sigma_\mathrm{sc}$) in the FC mode is almost constant at low temperatures and can be well modeled considering a fully gapped isotropic $s$-wave superconducting order parameter, with 2$\Delta$/k$_\mathrm{B}$T$_\mathrm{C} = 3.84$, close to 3.53, the value expected for BCS superconductors. The London penetration depth, superconducting carrier density and its effective mass are also estimated from TF-$\mu$SR analysis. Furthermore, zero field $\mu$SR confirms that there is no spontaneous magnetic moment, thus demonstrating that time-reversal symmetry is preserved in the superconducting ground state. Additionally, we showed through DFT calculations that Zr$_5$Pt$_3$ is a topological Dirac nodal-line semimetal protected by 6$_3$ screw rotations and glide mirror planes of the $P6_{3}/mcm$ space group. Therefore, this work puts forward a large, unprecedented class of superconducting topological nodal-line semimetals to realize novel quantum states of matter.

\paragraph{}

\begin{center} 
\bf Acknowledgments\\
\end{center}

\noindent PPF, LTFE and AJSM gratefully acknowledge the financial support of the São Paulo Research Foundation (FAPESP) under Grants 2019/05005-7, 2019/14359-7, and 2020/08258-0. KP acknowledge the financial support from DST India, for Inspire Fellowship (IF170620). AB would like to acknowledge DST India, for Inspire Faculty Research Grant (DST/INSPIRE/04/2015/000169), the SERB, India for core research grant support. and UK-India Newton funding for funding support. DTA would like to thank the Royal Society of London for Newton Advanced Fellowship funding and International Exchange funding between UK and Japan. The research was partially carried out using high-performance computing resources made available by the Superintend\^encia de Tecnologia da Informa\c c\~ao (STI), Universidade de S\~ao Paulo. The authors also acknowledge the National Laboratory for Scientific Computing (LNCC/MCTI, Brazil) for providing HPC resources of the SDumont supercomputer, which have contributed to the research results reported within this paper. This study was financed in part by the Coordenação de Aperfeiçoamento de Pessoal de Nível Superior – Brasil (CAPES) – Finance Code 001.

\section*{References}
\bibliography{refs}

\end{document}